\documentclass{aa}  

\usepackage{graphicx}
\usepackage{txfonts}
\usepackage[colorlinks]{hyperref}
\def\aap{A\&A}
\def\apj{ApJ}
\def\apjs{ApJS}

\def \hi {\ion{H}{i}}
\def\h2{H$_2$}

\def\kms{km\,s$^{-1}$}

\def\deg{\hbox{$^\circ$}}
\def\arcmin{\hbox{$^\prime$}}
\def\arcsec{\hbox{$^{\prime\prime}$}}

\defcitealias{Kalberla2021}{Paper~I}
\defcitealias{Kalberla2024}{Paper~II}
\defcitealias{Kalberla2025}{Paper~III}

\begin{document}

\title{Molecular hydrogen in filaments at high Galactic latitudes}

\subtitle{}

   \author{P.\ M.\ W.\ Kalberla }

\institute{Argelander-Institut f\"ur Astronomie, University of Bonn,
           Auf dem H\"ugel 71, 53121 Bonn, Germany \\
           \email{pkalberla@astro.uni-bonn.de}
 }

   \authorrunning{P.\,M.\,W. Kalberla } 

   \titlerunning{Molecular hydrogen in filaments }

   \date{Received 12 March 2025 / Accepted 21 November 2025 }

  \abstract 
{Neutral atomic hydrogen (\hi) absorption lines can be used to probe the
  cold neutral medium (CNM) at high Galactic latitudes. Cold \hi\ with
  a significant optical depth from the GASKAP-\hi\ survey is found to be
  located predominantly if not exclusively within filamentary
  structures that can be identified as caustics with the Hessian
  operator. Most of these \hi\ filaments (57\%) are also observable in
  the far-infrared (FIR) and trace the orientation of magnetic field
  lines.  }
{We considered whether molecular hydrogen (\h2) might also be 
  preferentially associated with CNM filaments. }
{We analyzed 241 \h2\ absorption lines against stars and determined whether
  the lines of sight intersected \hi\ or FIR filaments. Using {\it Far
    Ultraviolet Spectroscopic Explorer (FUSE)} \h2\ data in the velocity
  range $ -50 < v_{\mathrm{LSR}} < 50 $ \kms, we traced 65 additional
  \h2\ lines for filamentary \hi\ and FIR structures in velocity and
  probed the \h2\ absorption for coincidences in position and
  velocity. }
{For 305 out of 306 positions, the lines of sight with
  \h2\ absorption intersect \hi\ filaments. In 120 cases, there is also
  evidence for a correlation with dusty FIR filaments. All of the 65
  available sight lines with known velocities intersect
  \hi\ filaments. In 64 cases, the \h2\ velocities are consistent with
  \hi\ filament velocities. For FIR filaments, an agreement is found for only 13 out
  of 14 \h2\ absorption lines.  }
{For the majority of \h2\ absorption lines, there is evidence that
  \h2\ is associated with cold \hi\ filaments.  Evidence of an
  association with FIR filaments is less compelling. Confusion along the
  line of sight limits the detectability of FIR filaments. For a
  comparable degree of UV excitation in the disk and lower Galactic halo,
  the formation rate of \h2\ appears to be enhanced in \hi\ filaments
  with increased CNM densities.  }


  \maketitle

\section{Introduction}
\label{Intro}

Neutral atomic hydrogen (\hi) gas is one of the most important tracers
of the structure and dynamics of the interstellar medium (ISM). Molecular
hydrogen (\h2) likewise is the simplest and most abundant molecule, and it
provides a link to star formation at different scales. We intend
to discuss the properties of the  cold neutral medium (CNM), in
particular, the transition between the atomic and molecular phase in the
diffuse atomic medium. This phase is characterized by typical
\hi\ excitation temperatures $ 30 \la T_{\mathrm{ex}} \la 100 $ K,
densities of $ 10 \la n_{\mathrm{HI}} \la 100 $ cm$^{-3}$, and dust
temperatures of $ 15 \la T_{\mathrm{d}} \la 20 $ K. The \hi\ column
densities are below $N_{\mathrm{H}} = 10^{21.7} $ cm$^{-3}$ and the
molecular gas fraction is low, typically $ f_{\mathrm{H2}} = 2
n_{\mathrm{H2}} /n_{\mathrm{H}} \la 0.1$ \citep{Snow2006,Wakelam2017}.

Detailed \hi\ observations in this temperature regime demand sensitive
optical depth measurements, which are only possible against sufficiently
strong continuum background sources. Within the past two decades, 372
unique \hi\ lines of sight, distributed throughout the sky, have been observed
by various authors with several instruments. A compilation of these
21 cm \hi\ absorption and emission data, called BIGHICAT, are summarized in the Supplemental
Table 1 of \citet{Naomi2023}. Significant absorption is listed at 306 BIGHICAT
positions. Yet another 462 positions became available only recently from
the Galactic Australian Square
Kilometre Array Pathfinder Pilot Phase II Magellanic Cloud \hi\ foreground observations
\citep[GASKAP-\hi,][]{Nguyen2024}. They cover an area of 250 square degrees
of the Milky Way foreground toward the Magellanic Clouds. In total, 2 714
positions have been observed, resulting in 691 absorption line
detections.

The balance of heating and cooling processes in the ISM results in a
multiphase medium. The CNM is in pressure balance with a warm neutral
medium (WNM) \citep{Wolfire2003} that dominates most of the observable
\hi\ emission. Within the past decade, the evidence mounted that structures on arcminute scales are shaped in a filamentary way and that \hi\ and far-infrared
(FIR) emission agree very well (e.g., \citealt{Clark2014} and
\citealt{Kalberla2016}). The FIR filaments considered here are always
associated with \hi. These FIR structures are also associated with
cold dust, are stretched out along the magnetic field lines, and have been
further investigated by \citet{Clark2019}, \citet{Peek2019},
\citet{Clark2019b}, \citet{Murray2020}, \citet{Kalberla2020},
\citet{Kalberla2023}, and \citet{Lei2024}.

Filaments are defined by their particular topology as caustics that
correspond to singularities of gradient maps (e.g.,
\citealt{Castrigiano2004} or \citealt{Arnold1985}). As a standard tool
for classifying these structures, \citet{Kalberla2021} (hereafter Paper I)
have used the Hessian operator, which is based on partial derivatives of
the intensity distribution. Filaments can be described in this way
as caustics that are associated with cold CNM, an
increased CNM fraction, and an enhanced FIR emissivity, and they are understood as
coherent \hi\ fibers with local density enhancements in
position-velocity space. \citet{Kalberla2024} (hereafter Paper II) has
shown that BIGHICAT \hi\ absorption components are exclusively located
in filaments with FIR counterparts. Moreover, \citet{Kalberla2025}
(hereafter Paper III) has demonstrated that 57\% of the
GASKAP-\hi\ absorption components are associated with FIR filaments. All
of these 691 CNM clouds with detectable absorption are located in
\hi\ filaments.  Velocities along \hi\ filaments are observable in
projection in the plane of the sky and are affected by turbulent motions
with typical dispersions $ 2.48 < \sigma v_\mathrm{turb} < 3.9 $
\kms. In a similar way, velocities of embedded small-scale structures in
optical depth are affected by turbulence. The derived scale-dependent
velocity distribution agrees with the first law by
\citet{Larson1979}.

\h2\ is the simplest and most abundant molecule in the ISM, and its
formation precedes the formation of other molecules
\citep{Wakelam2017}. In the face of the finding that the cold \hi\ is
predominantly organized in filaments, we consider here the question
whether \h2\ as the most significant molecular component in the
CNM at high Galactic latitudes might also be distributed in filaments. In
Sect. \ref{Obs} we analyze the available observations. The conclusions are drawn
in Sect. \ref{Discussion}, and a summary is given in Sect. \ref{Summary}.

\section{Observations and data reduction}
\label{Obs}

\subsection{Identifying \hi\ and FIR filaments}
\label{Filaments}

Filaments are thin, thread-like structures that can numerically be
characterized as caustics or local singularities in the data
distribution. In the following, we exclusively use the term filaments for
structures that satisfy this mathematical classification. Caustics in
two-dimensional intensity distributions are uniquely defined
(e.g., \citealt{Arnold1985} or \citealt{Castrigiano2004}). The Hessian matrix is used in elementary calculus to
determine the properties of singularities (for a detailed discussion, we
refer to \citetalias{Kalberla2024}, and we only present a brief
summary here).

  Caustics in the intensity distribution can
  be identified as critical points with eigenvalues $\lambda_- < 0
  $. Hessian eigenvalues and the associated eigenvectors were
  determined throughout the sky for the HI4PI survey \citep{HI4PI2016} in the
  velocity range $ -50 < v_{\mathrm{LSR}} < 50 $ \kms\ and simultaneously
  for the intensity distribution of {\it Planck} data at 857 GHz
  \citep{Planck2020}. To obtain an identical resolution for the two
  databases, the FIR data were smoothed and converted into an nside =
  1024 HEALPix grid. 

  The filaments in \hi\ and FIR were derived from the eigenvalue
  distributions\footnote[1]{For downloads of Hessian eigenvalue spectra
    and filaments, see
    \url{https://www.astro.uni-bonn.de/hisurvey/AllSky_gauss/index.php}}. The
  velocities $ v_\mathrm{HI} $ along an \hi\ filament are the velocities
  of the local minima of the \hi\ eigenvalues $\lambda_-(v) $ (for
  examples, see Fig. A.1 of \citetalias{Kalberla2025}).  All FIR filaments
  were checked for coincidences with \hi\ filaments. To fit
  FIR filaments that are aligned with 
  \hi\ structures, the orientation angles of the Hessian eigenvectors
  were used. The velocity at which the best alignment between \hi\ and FIR
  orientation angles is found defines the FIR filament velocity
  $v_\mathrm{FIR}$. A tight agreement between FIR and \hi\ filaments was
  found in narrow velocity intervals of 1 \kms\ \citepalias[see Table 1
    in][]{Kalberla2021}. The correlation degrades significantly for
  larger velocity intervals with the implication that the \hi\ that is
  observed in filaments and is associated with the 857 GHz FIR structures
  has to be cold.  Using GASKAP-\hi\ absorption data observed by
  \citep{Nguyen2024},  \citetalias{Kalberla2025} verified that
  these filaments contain CNM.

  The correlation analysis is significantly complicated by the
  fact that the line of sight may intersect several filaments. For \hi\, these filaments are separated in velocity, and no confusion affects the derived parameters. For the FIR, however, the
  contributions from different filaments (indicated by \hi\ structures)
  and other background sources cannot be separated. Only the strongest
  FIR filaments (with low confusion) can be correlated with \hi.
  \citetalias{Kalberla2021} considered fiducial eigenvalues $\lambda_- < -1.5 $ K
  deg$^{-2}$ in FIR and $\lambda_- < -50 $ K deg$^{-2}$ in
  \hi\  to ensure a good correlation between
  \hi\ and FIR filaments. For these parameters, the average velocity
  uncertainties between \hi\ and associated FIR filaments are 2.6
  \kms. This rms scatter is associated with uncertainties in matching the
  \hi\ and FIR orientation angles. These differences are
  within 3\degr\ to 4\degr\ on average (fitting Voigt functions or Gaussians,
  respectively). The resulting velocity field along the FIR filaments
  projected onto the plane of the sky is homogeneous, with typical velocity
  fluctuations of 3.8 \kms.
  
\begin{figure}[ht] 
   \centering
   \includegraphics[width=8.5cm]{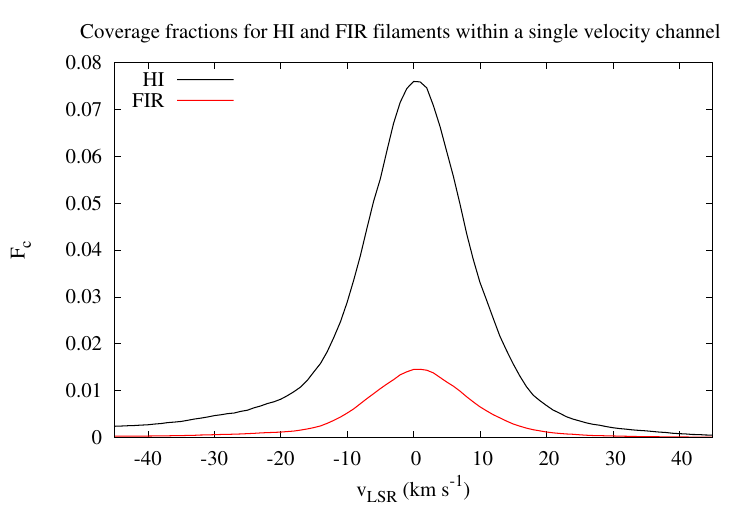}
   \caption{Average filament coverage fractions $F_c$ for FIR and \hi\ 
     filaments within a single velocity channel at Galactic
     latitudes $|b| > 10\degr$. }
   \label{coverage}
\end{figure}

  Considering the high-latitude sky with $|b| > 10\degr$ (or $|b| >
  30\degr$), we found that 30\% (18\%) of the observed nside=1024 HEALPix
  positions are covered by FIR filaments and 69\% (54\%) by
  \hi\ filaments.  Figure \ref{coverage} shows the covering fractions
  $F_c$ of filaments within single channels at $|b| > 10\degr$. The
  imbalance between the total number of \hi\ and FIR filaments is
  obvious: About five times more filaments are counted in \hi. Line of
  sight effects severely limitat the detection of the FIR
  filaments. A similar degradation is noticeable in \hi\ when
  the velocity resolution is decreased \citepalias{Kalberla2021}.  The curves shown
  in Fig.  \ref{coverage} are only estimates because they depend on
  constraints on the eigenvalues. Fig. A.2 in \citetalias{Kalberla2025}
  indicates that releasing the constraint $\lambda_- < -50 $ K
  deg$^{-2}$ for \hi\ filaments can lead to a detection rate that is higher by 21\% when it is applied to GASKAP-\hi\ absorption components.

  The global distribution of \hi\ and FIR filaments at high Galactic
  latitudes shown in Fig. \ref{coverage} has a dispersion of $8.1 \pm
  0.1$ \kms\ and $7.9 \pm 0.1$ \kms. All of the filaments have
  narrow line widths. They represent a population of small-scale CNM
  clumps that cannot be resolved individually in observations. These
  distributions therefore represent global probability distributions for
  the velocity of CNM clumps in the local vicinity.
  
\subsection{Stellar \h2\ absorption data, and unknown \h2\ velocities}
\label{Stellar}

\citet{Gudennavar2012} collated absorption line data toward 3008
stars in a unified database of interstellar column densities toward
stellar background sources. We used 419 entries for \h2\ column densities
in their Table 2 from the last published values. In total, 416 of these lines
of sight intersect \hi filaments, and 221 of these also intersect FIR filaments. Sources in
the Galactic plane might be affected by confusion, and we therefore restricted
our analysis in all cases to positions at latitudes $|b| > 10
\deg$. This sample contains 229 sources, and the lines of sight of 228 sources intersect
\hi\ filaments, and 95 lines of sight intersect FIR filaments. We conclude that 99.5\% of all
detected \h2\ lines show evidence for a correlation with \hi\ filaments.
For 41.7\% of the lines of sight, there is also evidence of a
correlation with dusty FIR filaments. We conclude that for the sample
considered in this section, the detection rate within the filaments increases
by 40\% with respect to a random distribution.  As discussed in
Sect. \ref{Filaments}, the identification of FIR filaments might be
limited by confusion. A low identification rate for FIR filaments
implies that most of the FIR filaments are weak, with less significant
Hessians.  In translucent sight lines, \citet{Rachford2002} and
\citet{Rachford2009} observed 12 high-latitude lines of sight. All of
them intersected \hi\,, and 8 lines of sight (66.7\%) intersected FIR filaments, indicating more prominent and better defined FIR filaments in this case.

\subsection{{\it FUSE} \h2\ absorption data, and known \h2\ velocities}
\label{FUSE}

The tabulated \h2\ data we discussed in the previous subsection do not
contain any information on the velocity centroids $v_\mathrm{H2}$ of the
\h2\ components. We discuss \h2\ data from the {\it Far Ultraviolet
  Spectroscopic Explorer (FUSE)} with known velocities that can be
compared with the velocities $v_\mathrm{fil}^{\mathrm{HI}} $ and
$v_\mathrm{fil}^{\mathrm{FIR}} $ of \hi\ and FIR filaments below.

A comprehensive database was provided by \citet{Wakker2006}, and we used
65 entries that are available within the velocity range $ -50 <
v_{\mathrm{LSR}} < 50 $ \kms.\footnote[2]{\citet{Wakker2006} presents
  measurements of \h2\ column densities toward 73 extragalactic targets
  but discusses predominantly absorption within high and
  intermediate velocity clouds.} For each observed line of sight, we
first verified whether it was located along a known FIR filament and then
derived the velocity difference $ \Delta v_\mathrm{FIR} = v_\mathrm{H2}
- v_\mathrm{fil}^{\mathrm{FIR}} $ for the associated FIR filament. Next,
we considered \hi\ filaments with the velocity difference $ \Delta
v_\mathrm{HI} = v_\mathrm{H2} - v_\mathrm{fil}^{\mathrm{HI}} $. For multiple \hi\ filaments, the closest match in velocity was selected. As
discussed by \citetalias{Kalberla2025}, these velocity differences
represent local turbulent velocity deviations between the \h2, which are probed by
absorption within a pencil beam, and the FIR filament with a typical
width of 9\arcmin. This corresponds to 0.63 pc at an assumed filament
distance of 250 pc. For the \hi\ absorption discussed by
\citetalias{Kalberla2025}, dispersions of $\sigma
v_\mathrm{turb}^{\mathrm{HI}} = 2.48 $ \kms\ for the whole \hi\ sample,
but $\sigma v_\mathrm{turb}^{\mathrm{FIR}} = 3.9 $ \kms\ for FIR
filaments was found. We use $\sigma
v_\mathrm{turb}^{\mathrm{HI}} = 2.48 $ \kms\ below as a reference for
comparison.

\begin{figure}[ht] 
   \centering
   \includegraphics[width=8.5cm]{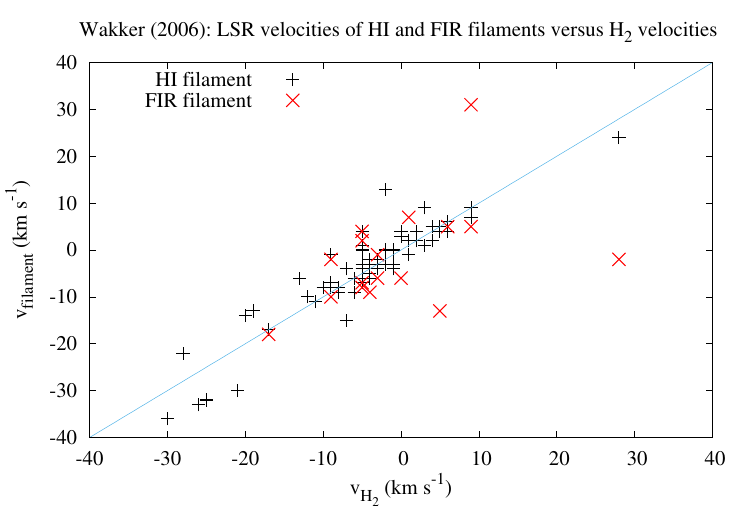}
   \includegraphics[width=8.5cm]{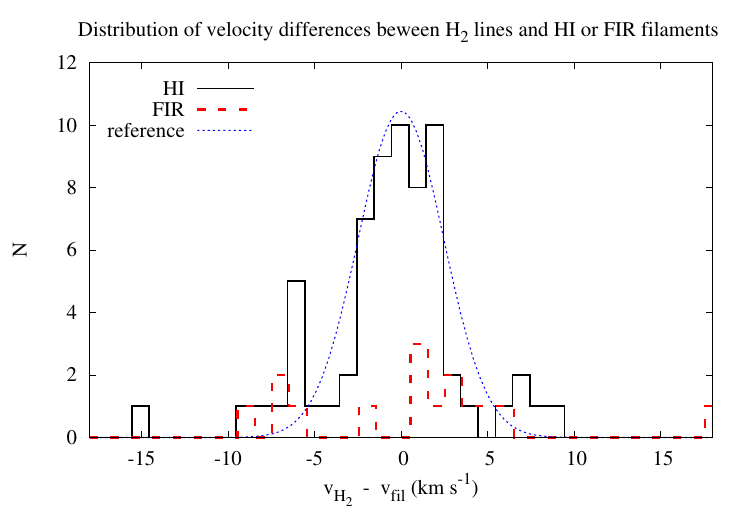}
   \caption{Comparison of velocities from \h2\ absorption and filaments
     in \hi\ and FIR. Top: Direct comparison between $ v_\mathrm{H2}$
     and filament velocities $ v_\mathrm{HI} $ (black) or $
     v_\mathrm{FIR} $ (red) as defined in Sect. \ref{Filaments}. Bottom:
     Histogram of the velocity deviations $ v_\mathrm{H2} -
     v_\mathrm{fil}^{\mathrm{HI}} $ for \hi\ and FIR
     filaments. \h2\ data are taken from \citet{Wakker2006}. The dotted blue
     curve represents a reference distribution with $\sigma
     v_\mathrm{turb}^{\mathrm{HI}} = 2.48 $ \kms. }
   \label{Wakker}
\end{figure}

We found that all of the 65 low-velocity \h2\ components from the
catalog of \citet{Wakker2006} are associated with \hi\ filaments, but only
14 are associated with FIR filaments. The results are displayed in Fig. \ref{Wakker}.
At the top, we show a comparison of \h2\ LSR velocities with the velocities
of FIR filaments. The black symbols show a nearly linear relation
between \h2\ and \hi\ velocities, and the deviations have a dispersion of
$\sigma_\mathrm{HI} = 4 $ \kms. At the bottom of Fig. \ref{Wakker}, we show
histograms of the velocity deviations $ v_\mathrm{H2} - v_\mathrm{fil}
$.  We compare these distributions with the reference distribution of
$\sigma \sim 2.48 $ \kms, as determined by \citetalias{Kalberla2025}.
The observed deviations $ \Delta v_\mathrm{HI} $ (black) are compatible with
this reference, except for a few outliers with $ |v_\mathrm{H2} -
v_\mathrm{fil}^{\mathrm{HI}}| \ga 10 $ \kms\ ($4 \sigma$ with
  respect to the reference distribution). The velocity deviations $ \Delta
v_\mathrm{FIR} $ (red) are barely compatible with the reference distribution.

\begin{figure}[ht] 
   \centering
   \includegraphics[width=8.5cm]{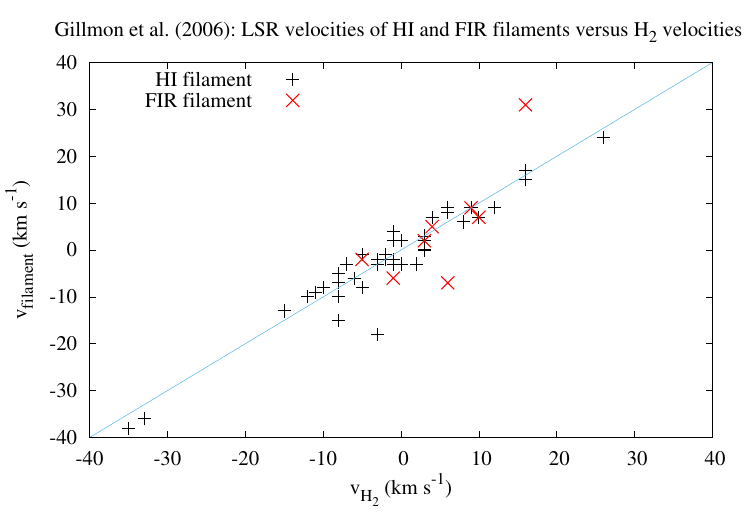}
   \includegraphics[width=8.5cm]{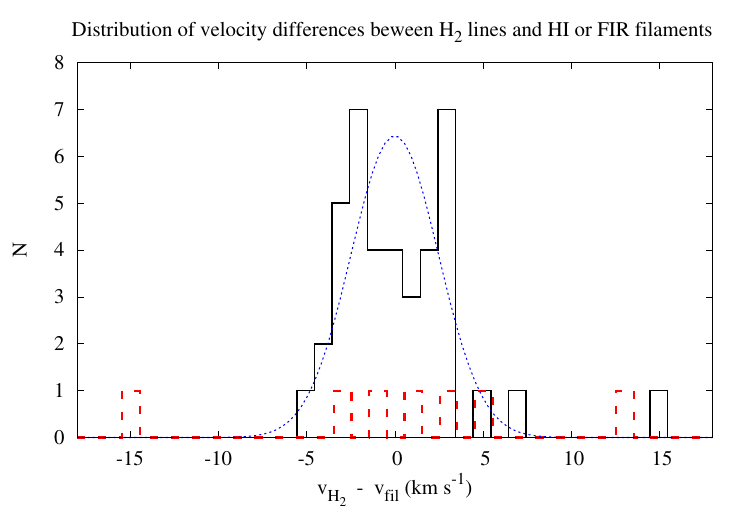}
   \caption{Comparison of velocities from \h2\ absorption and filaments
     in \hi\ and FIR. Top: Direct comparison between $ v_\mathrm{H2}$
     and filament velocities $ v_\mathrm{HI} $ (black) or $
     v_\mathrm{FIR} $ (red) as defined in Sect. \ref{Filaments}.
     Bottom: Histogram of the velocity deviations $ v_\mathrm{H2} -
     v_\mathrm{fil}^{\mathrm{HI}} $ for \hi\ and FIR
     filaments. \h2\ data are taken from \citet{Gillmon2006a}. The dotted blue
     curve represents a reference distribution with $\sigma
     v_\mathrm{turb}^{\mathrm{HI}} = 2.48 $ \kms. }
   \label{Gillmon}
\end{figure}

An alternative database with 40 {\it FUSE} \h2\ absorption lines was
provided by \citet{Gillmon2006a}. We processed these data in the same
way as the data by \citet{Wakker2006} discussed above. Figure
\ref{Gillmon} displays the results for 40 absorption lines. The plot at the
top shows a good linear relation between \h2\ and \hi\ velocities, in
this case, with a dispersion of $\sigma_\mathrm{HI} = 3.6 $ \kms. The
histogram for $ v_\mathrm{H2} - v_\mathrm{fil} $ (bottom) shows a
somewhat weird double-peaked structure, but has fewer outliers.  
  The details of the velocity distributions shown in Figs. \ref{Wakker} and
  \ref{Gillmon} might be affected by line blending, or by
  ambiguities in assigning \hi\ filaments to \h2\ absorption
  structures in a few cases. The general characteristics of the velocity distributions
  remain unaffected, however.  

The 40 sight lines used by \citet{Gillmon2006a} are in common with
\citet{Wakker2006}. The discrepancies between Figs. \ref {Wakker} and
\ref{Gillmon} can be explained by problems in the {\it FUSE} wavelength
calibration and different approaches to solving the calibration uncertainties.
The {\it FUSE} velocity resolution at the full width at half maximum (FWHM)
depends on the data binning and typically is 20
\kms\ \citep[][Sect. 4.2]{Wakker2006}. The velocity centroids were estimated
by \citet[][Sect. 2.2]{Gillmon2006a} to be uncertain by 5
\kms on average. Calibration problems and uncertainties of 2 \kms\ for sight lines
with a high signal-to-noise ratio (S/N) and simple absorption-line
structure up to 6–8 \kms\ for low S/N data were reported by
\citet[][Sect. 2.2]{Wakker2006}\footnote[3]{\citet{Gillmon2006a} required
  S/R $> 4$, after binning 8 pixels at most.} (for details, we refer to the extended discussions in that paper). We derived average
velocity discrepancies with a dispersion of 5.7 \kms\ from an
intercomparison of the two datasets that were used to derive
Figs. \ref{Wakker} and \ref{Gillmon}. Typical 21 cm emission lines in
cold \hi\ filaments have dispersions of $\sim 1.3$
\kms\ (e.g., \citealt{Clark2014} and \citealt{Kalberla2016}). Filaments,
using the Hessian operator, were determined on a velocity grid with a resolution of
1 \kms\ \citepalias{Kalberla2021}.

  All of the velocity uncertainties discussed above were
  compared with observed turbulent velocity fluctuations for the CNM of about $ \sigma_v = 2.48 $ \kms\ on arcminute scales in the plane
  of the sky and within a single-dish beam \citepalias{Kalberla2025}. For prominent FIR filaments, the observed dispersion can be up to
  3.9 \kms. According to the first law by \citet{Larson1979}, turbulent
  motions cause scale-dependent velocity fluctuations in the ISM. The
  velocity dispersion increases with distance $l$ as $\sigma_\mathrm{v}
  (l) \propto l^q$. For supersonic turbulence, the exponent $q$ is
  expected to be in the range $1/3 \la q \la 1/2$; $q = 0.5$ applies to
  the molecular cloud regime \citep{Heyer2009} and also to the CNM. The
  relation of line width to size was verified for \hi\ absorption components
  over distances from 7\arcsec to 10\degr \citepalias[][Fig. 4
  ]{Kalberla2025}.
  
To gain a deeper insight into the properties of the \h2\ distribution, we
plot in Fig. \ref{fH2} the \h2\ fraction $ \mathrm{log}f_\mathrm{H2} $
derived by \citet{Gillmon2006a} as a function of $ v_\mathrm{H2} -
v_\mathrm{fil}$. Figure \ref{NH2} alternatively shows the relation
between \h2\ column densities and $ v_\mathrm{H2} - v_\mathrm{fil}$
derived by \citet{Wakker2006}.  As noted before by \citet{Gillmon2006a},
$ \mathrm{log}N_\mathrm{H2} \sim 17 $ appears to separate two
populations of \h2\ absorbers. A similar gap exists in Fig. \ref{fH2} at
$ \mathrm{log}f_\mathrm{H2} \sim -3 $. Adopting turbulent velocity
fluctuations with $ \sigma_\mathrm{turb} = 2.48 $ \kms\ as a reference,
we considered $|v_\mathrm{H2} - v_\mathrm{fil}| \ga 10 $ \kms\ as
misidentifications and excluded these positions from the
discussion. Because of the low number statistics for FIR filaments, we also
excluded FIR filaments. With these restrictions, the scatter
$|v_\mathrm{H2} - v_\mathrm{fil}|$ increases in Fig \ref{NH2} for $
\mathrm{log}N_\mathrm{H2} \la 17 $ by a factor of two with respect to $
\mathrm{log}N_\mathrm{H2} \ga 17 $. The average dispersion is 3.6
\kms\ and in the range $2.48 < \sigma v_\mathrm{turb } < 3.9 $ \kms,
observed for \hi\ absorption discussed by \citetalias{Kalberla2025}.
Turbulence is also observable in the distribution of the metal lines.
Comparing \h2\ velocities with the velocities of metal lines,
\citet{Wakker2006} derived from his Table 2 a compatible dispersion of
3.9 \kms\ with peak deviations up to $\pm 9$ \kms.  For the
wavelength calibration problems discussed by \citet[][Sect. 2.2
    and 4.2]{Wakker2006},  it is plausible that the center velocities
of sight lines with a low column density have significantly increased
uncertainties. These uncertainties also explain the differences in
  the dispersions that are visible in Figs. \ref{Wakker} and
  \ref{Gillmon}. The extended wings in Figs. \ref{Wakker} are caused by
  sources with low \h2\ column densities, which explains the high uncertainties in
  the velocity centroids.  In the light of these problems, we conclude
that the observed scatter in $ v_\mathrm{H2} - v_\mathrm{HI}$ and $
v_\mathrm{H2} - v_\mathrm{FIR}$ is significantly affected by
  observational uncertainties, but is still consistent with the turbulent
velocity field shown in Fig. 4 of \citetalias{Kalberla2025}.

\section{Discussion}
\label{Discussion}

The ISM is dominated by a multiphase medium \citep{Wolfire2003} with a
diffuse WNM and a clumpy CNM in pressure equilibrium. Large-scale
\hi\ surveys such as HI4PI \citep{HI4PI2016} or GALFA-\hi\ \citep{Peek2018}
have led to the picture that a major part of the cold ISM is organized
in filamentary structures that are aligned with the magnetic field lines
(e.g., \citealt{Clark2014} and \citealt{Kalberla2016}). These structures
are detectable with the Hessian operator as caustics. Based on this
classification, filaments can rigorously be characterized in the FIR with a
coherent velocity field and in \hi\ with an extended network of
\hi\ fibers at distinct velocities \citepalias{Kalberla2024}. 
\hi\ and FIR filaments are both exposed to a turbulent external medium,
which leads to characteristic velocity fluctuations along and within the
filaments.

A detailed analysis of the CNM in filaments became available with the
GASKAP-\hi\ absorption line survey, which covers nine adjacent fields of with over 25 square-degrees in the direction to the Magellanic clouds
\citep{Nguyen2024}. The analysis of 691 absorption components in 462
lines of sight has led to major improvements in the understanding of the
physical properties of the CNM. It was subsequently found that all
\hi\ absorption line components are associated with filamentary
structures in \hi\ and/or FIR \citepalias{Kalberla2025}, suggesting that
a major part of the CNM is located in filaments.  The
  GASKAP-\hi\ components have a mean column density of $ N_\mathrm{HI} =
  1.8~10^{20} \mathrm{cm}^{-2}$ and a mean spin temperature of $ T_s = 50$
  K, and in pressure eqilibrium with the WNM with $\log (p/{\mathrm{k}})
  = 3.58$ \citep{Jenkins2011}, a mean volume density $ n_\mathrm{HI} =
  76~ \mathrm{cm}^{-3}$ can be derived. The question arises whether these conditions support a transition from \hi\ to \h2. If
  this is the case, we expect that \h2\ might also share the filamentary
  nature of the CNM.

\begin{figure}[ht] 
   \centering
   \includegraphics[width=8.5cm]{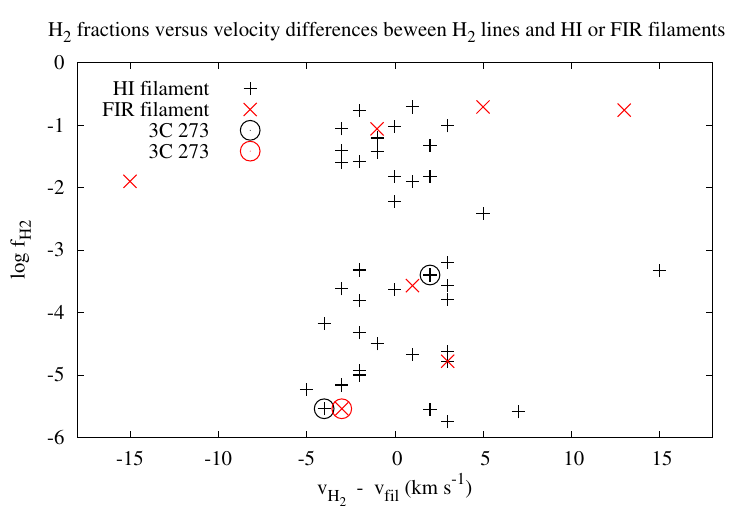}
   \caption{\h2\ fraction $ \mathrm{log}f_\mathrm{H2} $ derived by
     \citet{Gillmon2006a} vs. observed velocity deviations $
     v_\mathrm{H2} - v_\mathrm{fil}$. The filament velocities are $v_\mathrm{HI}
     $ (black) or $ v_\mathrm{FIR} $ (red), as defined in
     Sect. \ref{Filaments}. 
   }
   \label{fH2}
\end{figure}

\begin{figure}[ht] 
   \centering
   \includegraphics[width=8.5cm]{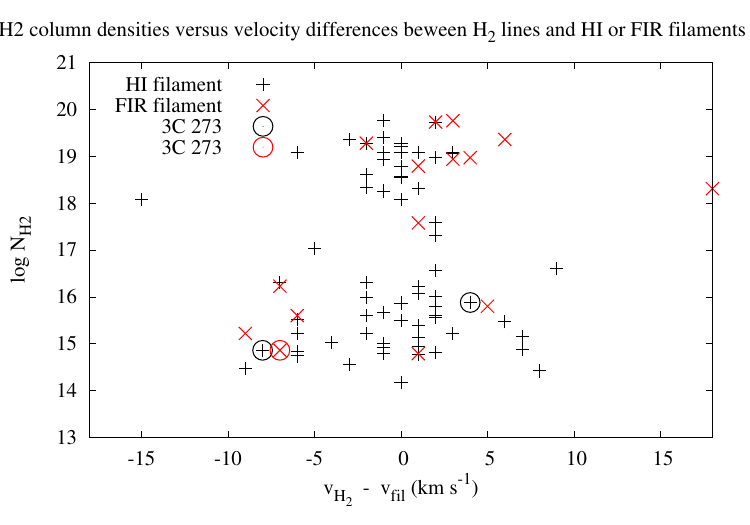}
   \caption{\h2\ column densities $ \mathrm{log}N_\mathrm{H2} $ derived
     by \citet{Wakker2006} vs. observed velocity deviations $
     v_\mathrm{H2} - v_\mathrm{fil}$. The filament velocities are $v_\mathrm{HI}
     $ (black) or $ v_\mathrm{FIR} $ (red), as defined in
     Sect. \ref{Filaments}. }
   \label{NH2}
\end{figure}

Analyzing published \h2\ absorption data, we found that 305 out
of 306 lines of sight are located along \hi\ filaments in total. A subset of 64
absorption lines has known \h2\ center velocities. These components,
again with a single exception, can also be assigned to \hi\ filaments in
velocity.\footnote[4]{For seven sources observed by \citet{Wakker2006}
  no significant \h2 lines could be detected. Velocities
  determined from metal lines were found in four of these cases to be
  consistent with \hi\ filament velocities, however. For a single source, no caustic
  was found, and two cases remained without conclusive absorption lines.} The observed
deviations $ v_\mathrm{H2} - v_\mathrm{HI} $ are consistent with the
turbulent velocity fluctuations from the GASKAP-\hi\ data
\citepalias{Kalberla2025}.  Assuming that the \h2\ formation is
supported in presence of CNM with parameters as observed by GASKAP-\hi,
we conclude that the observable \h2\ may predominantly be located
in cold \hi\ filaments.

\begin{figure}[ht] 
   \centering
   \includegraphics[width=8.5cm]{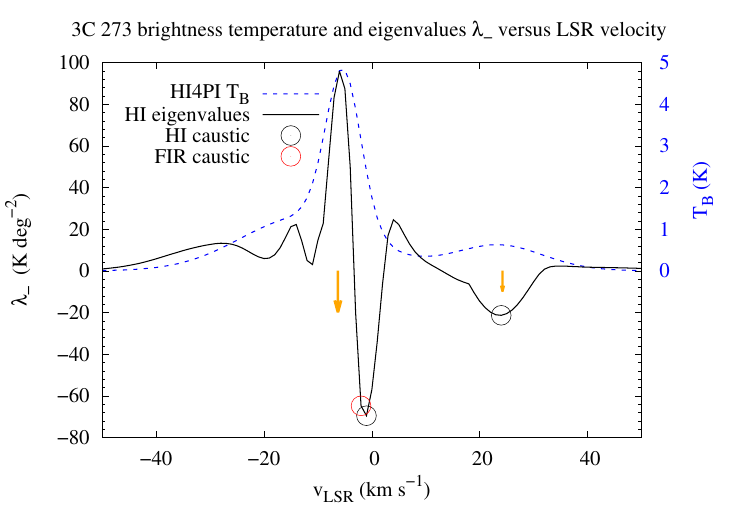}
   \caption{HI4PI brightness temperatures (blue) at the position of 3C
     273 and associated eigenvalues $\lambda_- $ (black).  \hi\ caustics
     exist at $v_{\mathrm{LSR}} = -1$ and + 24 \kms.  The FIR caustic is
     at $v_{\mathrm{LSR}} = -2$ \kms. The arrows sketch absorption
     components observed by \citet{Murray2018} and \citet{Heiles2003} at
      $v_{\mathrm{LSR}} = -6.3$ and  $v_{\mathrm{LSR}} = 24.2$ \kms. }
   \label{3c273_lam1}
\end{figure}

This inference is indirect and needs \hi\ absorption data for
verification. These observations are missing, except for a single
case. Only for the source 3C273 (QSO B1226+0219) can a link between \h2,
\hi, and dust be established. The line of sight to 3C273 intersects
an FIR filament with a dust temperature of $T_\mathrm{d} = 17.978
\pm 0.013$ K \citep{Shull2024}.  Filaments are observed for \hi\ at
$v_\mathrm{HI} = -1 $ \kms\ and for the FIR at $v_\mathrm{FIR} = -2 $
\kms (see Fig. \ref{3c273_lam1}).

In addition, an \hi\ filament lies at $v_\mathrm{HI} = 24
$ \kms. We discuss this structure first. A corresponding \h2\ absorption
was observed by \citet{Gillmon2006a} at $v_\mathrm{LSR} = 26 $ \kms\ and
by \citet{Wakker2006} at $v_\mathrm{LSR} = 28 $ \kms\footnote[5]{In his
  Table 4 a velocity $v_\mathrm{LSR} = 25 $ \kms\ is given.}.  The
\hi\ absorption with $\tau = 0.002 $ is too weak for us to measure a
significant spin temperature, but \citet{Heiles2003} determined with the
Arecibo telescope an upper limit of 87 K for the excitation temperature
at this velocity. For a local equilibrium pressure of $\log
(p/{\mathrm{k}}) = 3.58$ \citep{Jenkins2011}, we estimated the density
to be $n_\mathrm{CNM} \la 44 ~ \rm \mathrm{cm}^{-3}$.  This is
evidence that \h2\ at the velocity $24 \la v_\mathrm{LSR} \la 28 $
\kms\ is associated with an \hi\ filament at $v_\mathrm{LSR} = 24 $
\kms.

Particular cold \hi\ was observed at $v_\mathrm{LSR} = -6.3 $
\kms. \citet{Heiles2003} determined a spin temperature of $T_\mathrm{s}
= 44.4 $ K, and \citet{Murray2018} reported $T_\mathrm{s} = 17 $ K. In
pressure equilibrium, the density is $86 \la n_\mathrm{CNM} \la 223 ~
\rm \mathrm{cm}^{-3}$.  This component is associated with \hi\ gas at $
v_\mathrm{LSR} = -5.5 $ \kms\ \citep{Heiles2003} or $ v_\mathrm{LSR} =
-5.8 $ \kms\ \citep{Murray2018}, belonging to the unstable neutral
medium (UNM) with a spin temperature of 651 K or 455 K, respectively. As
discussed by \citet[][Sect. 4.5.2 ]{Heiles2003}, overlapping opacity
components cause considerable uncertainties in the Gaussian fit
parameters.  The \hi\ velocity of $v_\mathrm{LSR} = -6.3 $ \kms\ is well
defined and agrees with the \h2\ absorption observed at $v_\mathrm{H2} =
-5 $ \kms\ by \citep{Gillmon2006a}, but differs significantly from
$v_\mathrm{H2} = -9 $ \kms\ derived by \citet[][Sect. 4.2]{Wakker2006}
in presence of considerable uncertainties in the \h2\ velocity
calibration. The observations by \citet{Heiles2003} indicated CNM column
densities of $ N_\mathrm{HI} = 4~10^{18} \mathrm{cm}^{-2}$ with
associated column densities of $ N_\mathrm{HI} = 46~10^{18}
\mathrm{cm}^{-2}$ for the UNM. \citet{Murray2018} determined for the CNM
$ N_\mathrm{HI} = 2~10^{18} \mathrm{cm}^{-2}$, and for the UNM, $
N_\mathrm{HI} = 29~10^{18} \mathrm{cm}^{-2}$. Despite the observational
uncertainties, we conclude that the structures observed in FIR, \h2\ and
\hi\, are related. 3C273 is one of the strongest continuum sources,
allowing high S/N detections in \hi\ and \h2\ absorption.  The velocity
deviations $ v_\mathrm{H2} - v_\mathrm{HI} $ are within the
uncertainties that are consistent with the turbulent velocity field from
GASKAP-\hi\ \citepalias{Kalberla2025}. The filaments are indicated in
Fig. \ref{NH2}.

Consistent with previous GASKAP-\hi\ results, our current analysis leads
to detection rates close to 100\% for absorption lines associated with
\hi\ filaments. We compare this with expectations for a random
filament distribution. The joint probability $P$ to detect absorption along
a filament depends on $F_c$, the average filament coverage per channel, as shown
in Fig. \ref{coverage}, on the filament velocity $ v_\mathrm{fil}$, and
further, on the expected turbulent velocity dispersion 
$\sigma_\mathrm{turb}$ along the filament,
\begin{equation}
     P  = \int{ F_c(v) \frac{1}{\sqrt{2\pi\sigma_\mathrm{turb}^2}}
       \mathrm{exp}^{-\frac{(v-v_\mathrm{fil})^2}{2\sigma_\mathrm{turb}^2}} dv }.
  \label{eq:P}
\end{equation}
Using $\sigma_\mathrm{turb}^{\mathrm{HI}} = 2.48 $ \kms\ and
$\sigma_\mathrm{turb}^{\mathrm{FIR}} = 3.9 $ \kms\, we derived for
\hi\ filaments an expected probability of $ P_\mathrm{HI} = 0.112 \pm
0.033 $ and $ P_\mathrm{FIR} = 0.0085 \pm 0.0037 $ for FIR
filaments. For reference, we compared these results with
GASKAP-\hi\ absorption observations \citep{Nguyen2024}. In this case, we
derived probabilities of $ P_\mathrm{HI} = 0.126 \pm 0.013$ for \hi\ and $ P_\mathrm{FIR} = 0.0112 \pm 0.0014$ for the FIR filaments
discussed by \citetalias{Kalberla2025}. In summary, compared to a random
distribution, the observed detection rate for \h2\ or \hi\ absorption
associated with \hi\ filaments is higher by a factor of eight to nine
and even higher than 20 for FIR filaments.

The derived overdensity suggests that the cold and dense regions that
give rise to detectable absorption lines in \h2\ and \hi\ are located
within \hi\ filaments.  The detection rate $R$ for FIR filaments,
however, is affected by systematical environmental fluctuations. For GASKAP-\hi\ data in the direction of two prominent FIR filaments, a rate
of $R = 0.57$ was found \citepalias{Kalberla2025}. For stellar
\h2\ lines, which are distributed throughout the high-latitude sky, we obtained $R =
0.43$. This might be representative for the high-latitude sky. For the
{\it FUSE} absorption observations lines against AGN, we only obtained $R =
0.26$. A sample in the direction of translucent sight lines yields
$R = 0.66$, however. The margin in $R$ can be explained by selection effects. We
considered the two extreme cases. The sample studied by
\citet{Wakker2006} is in directions with low extinction ($R = 0.26$),
while translucent sight lines ($R = 0.66$) contain significant amounts
of dust. These regions can provide sufficient protection from
interstellar radiation, which would support an increased
\h2\ content. Photodissociation is assumed to be generally the dominant
\h2\ removal mechanism \citep{Wakelam2017}, and less dissociation is
expected for translucent sight lines. Dissociation is also minimized for
condensed structures with high density. We therefore rejected the ideas that
absorption lines and filaments might be caused by an observational
accumulation from numerous separate volume elements along the line of
sight by a velocity-crowding effect \citep{Yuen2024}.

The case of \h2\,, which might be produced in structures that are much
denser than the clouds on average, was considered by
\citet{Valdivia2016}. These authors performed high-resolution
magnetohydrodynamical (MHD) colliding-flow simulations to study the
\h2\ formation of multiphase molecular clouds. As a result of a
combination of thermal pressure, ram pressure, and gravity, the clouds
produced at the converging point of HI streams were highly
inhomogeneous. These authors demonstrated in their Fig. 13 that the
derived molecular fractions as function of total hydrogen column
density compare well with observations by \citet{Gillmon2006a} and
\citet{Rachford2002,Rachford2009}. Their Fig. B.1 shows that the total
shielding coefficient for \h2\ at column densities
$\mathrm{log}N_\mathrm{H2} \la 20$ is little affected by an absence of
shielding by dust. The translucent sight lines observed by
\citet{Rachford2002,Rachford2009} have $\mathrm{log} N_\mathrm{H2} \ga
20 $; shielding is expected to be strong. Thus, environmental
conditions might explain our finding that the detection rate $R$ for
\h2\ absorption in FIR filaments is higher in translucent regions.

The MHD colliding-flow simulations by \citet[][Figs. 8 and
    9]{Valdivia2016} showed that \h2\ can be formed quickly, within two or
  three million years for typical volume densities and spin temperatures as
  observed in the GASKAP-\hi\ sample. Compressible forcing leads to the
  formation of clumps that are significantly denser than for
  solenoidal turbulent forcing \citep{Micic2012}. For 3C273,
  which lies in direction of radio loops I and IV
  \citep{Panopoulou2021} at a distance of $\sim 112$ pc, there is some
  evidence that the observed \hi\ distribution might be affected by shocks
  from supernova events. In comparison to the CNM, ten times more UNM is
  observed. Most of the cold \hi\ is therefore probably in transition from the
  WNM to the CNM.

  Processes that lead to an \hi\ to \h2\ transition for turbulent
  forcing under various conditions were studied with MHD simulations
  by \citet{Bellomi2020}. In this case, the observed correlation between
  the \hi\ and \h2\ column density \citep[][Fig. 3]{Bellomi2020} is
  reproduced well for \h2\,, which is built up in CNM structures between
  $\sim 3$ and $\sim 10$ pc. \h2\ for fractions $
  \mathrm{log}f_\mathrm{H2} \la -3 $ and column densities $
  \mathrm{log}N_\mathrm{H2} \la 17 $ are built up from diffuse low-density components along the line of sight, while dense small-scale
  clumps cause high \h2\ column densities
  \citep[][Fig. 7]{Bellomi2020}. The locations of the 3C273 data within
  Figs. \ref{fH2} and \ref{NH2} are indicated in both cases in the
  lower parts of the plots. This regime, based on the model by
  \citet{Bellomi2020}, is expected to be occupied by diffuse low-density
  components ($ n \la 8~ \mathrm{cm}^{-3}$). The observed CNM densities
  for 3C273 ($ n_\mathrm{CNM} \ga 86~ \mathrm{cm}^{-3}$) exceed the
  model distribution significantly. Likewise the mean densities from the
  GASKAP-\hi\ sample ($ n_\mathrm{CNM} \sim 76~ \mathrm{cm}^{-3}$) do
  not fit the model in Fig. 7 of \citet{Bellomi2020}. We conclude
  that a model for turbulent forcing without additional compressible
  contributions is not supported by \hi\ absorption
  observations so far.

\section{Summary and conclusions}
\label{Summary}

Recent sensitive \hi\ absorption observations \citep{Nguyen2024}
  have shown that the location of CNM in the plane of the sky is
  predominantly along \hi\ filaments \citepalias{Kalberla2025}, as
  observed with large single-dish telescopes throughout the sky. A
  fraction of these filaments can also be traced in the FIR. The turbulent velocity distribution is characteristic
  for the 3D distribution along the filaments, which is consistent with the first law by
  \citet{Larson1979}. Inspired by these results, we investigated
  whether the observed \h2\ distribution is also consistent with the
  \hi\ results.

  We found that most of the known \h2\ absorption lines are located
  along \hi\ filaments. Only for a few of these data were velocity
  centroids determined as well, and
  \hi\ absorption data are available for only a single source. This is clearly an observational
  deficit that limitats the credibility of the data
  analysis. Additional deficits are caused by problems in the wavelength
  or velocity calibration of the {\it FUSE}
  observations. \citet{Wakker2006} and \citet{Gillmon2006a} provided
  different solutions with an average velocity discrepancy of 5.7 \kms.
  These data need to be considered with care. In 64 out of 65 lines of
  sight are the \h2\ velocities with derived turbulent velocity fluctuations
still consistent with the \hi\ filament velocities
  discussed by \citetalias{Kalberla2025}, however.

  High-sensitivity
  \hi\ absorption lines are available in the direction to 3C273, which is one of the
  strongest radio sources. These lines confirm the expectation that
  \h2\ is correlated with the CNM along filaments in position and
  velocity. \h2\ formation is expected to be enhanced in the presence of
  cold and dense \hi\ \citep[e.g.][]{Valdivia2016}.  We are confident
  that these results can be generalized, and we expect  that \h2\ with
increased densities predominantly exists within CNM filaments. 
  Additional \hi\ observations at positions with known \h2\ absorbers
  are encouraged. 

The formation of \h2\ in the gas phase is not very relevant to
interstellar environments \citep{Vidali2013,Valdivia2016}. The
preference of \h2\ to accumulate in \hi\ filaments only implies that the
CNM in filaments provides perfect conditions to support
\h2\ formation. These conditions are given by cold \hi\ associated with
cold dust \citep[][Fig. 17]{Nguyen2024} that has settled in
\hi\ filaments.  The direct observational evidence for \h2\ in FIR
filaments is affected by sensitivity limitations (see, e.g., Fig. 4 of
\citet{Kalberla2016}). Only the most prominent FIR filaments are easily
observable. Assuming, on the other hand, an increased formation rate for
\h2\ in regions with cold \hi\ at high Galactic latitudes leads to
filamentary structures on large scales \citep[][Fig. 11]{Kalberla2016},
without the need to invoke a sophisticated Hessian analysis (see also
\citealt{Skalidis2024}).


\begin{acknowledgements}
I thank the second referee for constructive criticism that helped to
improve the paper.  HI4PI is based on observations with the 100-m
telescope of the MPIfR (Max-Planck- Institut für Radioastronomie) at
Effelsberg and Murriyang, the Parkes radio telescope, which is part of
the Australia Telescope National Facility (https://ror.org/05qajvd42)
which is funded by the Australian Government for operation as a National
Facility managed by CSIRO. This research has made use of NASA's
Astrophysics Data System.  Some of the results in this paper have been
derived using the HEALPix package.
\end{acknowledgements}

\end{document}